\documentclass[aps,pra,twocolumn,nofootinbib]{revtex4-1}
\usepackage{epsfig}
\usepackage[colorlinks,linkcolor=blue,anchorcolor=blue,citecolor=blue,urlcolor=blue,breaklinks=true]{hyperref}
\usepackage{graphics}
\usepackage{color}
\usepackage{amsmath}
\usepackage{bm}
\usepackage{subfig}
\usepackage{braket}             
\usepackage{amsfonts}

\begin{document}
\author{Roman Sahakyan$^{1}$}      
\author{Romik Sargsyan$^{1}$}
\author{Edgar Pogosyan$^{2}$}
\author{Karen Arzumanyan$^{1,4}$}
\author{Emil A. Gazazyan$^{3,4}$}
\email{emil@quopt.net}
\affiliation{$^{1}$ Russian-Armenian University, 123 Hovsep Emin street, Yerevan, 0051 Armenia}
\affiliation{$^{2}$ Sirius University, Russian Federation, Krasnodar region, Sirius Federal Territory, 354340, Olympic Ave., 1.}
\affiliation{$^{3}$ Institute for Physical Research, of the National Academy of Sciences of the Republic of Armenia, Ashtarak-2, 0203, Republic of Armenia}
\affiliation{$^{4}$ Institute for Informatics and Automation Problems, of the National Academy of Sciences of the Republic of Armenia, Yerevan, 0014, 1, P. Sevak str., Republic of Armenia}

\title{Automated Optimization of Laser Fields for Quantum State Manipulation}
\begin{abstract}
  A gradient-based optimization approach combined with automatic differentiation is employed to ensure high accuracy and scalability when working with high-dimensional 
  parameter spaces. Numerical simulations confirm the effectiveness of the proposed method: the population is reliably transferred to the target state with minimal 
  occupation of intermediate levels, while the control pulses remain smooth and physically implementable.

The developed framework serves as a universal and experimentally applicable tool for automated control pulse design in quantum systems. 
It is particularly useful in scenarios where analytical methods or manual parameter tuning—such as standard schemes like STIRAP—prove to be inefficient or inapplicable.

\end{abstract}
\maketitle
\section{Introduction}

The ability to coherently steer population transfer between quantum states is central to numerous applications in atomic, molecular, and optical physics. 
Among the most effective approaches to achieve this is \textit{Stimulated Raman Adiabatic Passage} (STIRAP), a technique that exploits adiabatic evolution 
under tailored laser fields to enable near-complete and lossless population transfer. Unlike conventional excitation schemes, STIRAP uses a counterintuitive 
pulse sequence to establish a coherent superposition---referred to as a dark state---that dynamically connects the initial and target states while circumventing 
the intermediate state, thereby avoiding detrimental spontaneous decay. Initially introduced for the study of chemical reaction dynamics, STIRAP has since evolved 
into a broadly used control method across many quantum systems \cite{Bergmann2015stirap}.

Recent developments have significantly extended the reach of STIRAP beyond its original domain. As summarized in a comprehensive review \cite{Bergmann2019}, 
the method has been successfully implemented in a variety of experimental platforms, including the creation of ultracold molecules, precision metrology, 
manipulation of photonic and magnonic excitations, and quantum information processing in solid-state and superconducting systems. Moreover, theoretical advancements 
such as fractional STIRAP and spatial adiabatic protocols have expanded its applicability to engineered quantum superpositions and wave transport phenomena. 
These advances confirm STIRAP's status as a cornerstone methodology in the ongoing pursuit of robust, high-fidelity quantum control.
  
A seminal review by Vitanov et al.~\cite{Vitanov2017} provides a comprehensive theoretical and experimental account of STIRAP
and its numerous extensions. The authors present a systematic analysis of various coupling configurations ($\Lambda$, $\Xi$, and tripod),
explicate the key principles underpinning adiabatic population transfer, and survey a broad spectrum of experimental implementations.
Of particular note is the emphasis placed on the method’s robustness to moderate fluctuations in control parameters, as well as its limitations
under non-adiabatic and dissipative conditions.

To aid in conceptual understanding, Shore \cite{Shore:17} introduced a graphical and pedagogical representation of STIRAP,
offering intuitive insights into its geometric structure and dynamical behaviour. This tutorial-style exposition proves particularly valuable for
visualising adiabatic state trajectories and for designing control pulses grounded in the logic of dark-state navigation.

The STIRAP method continues to evolve actively, with recent research focusing in particular on its modification through the use of
frequency-modulated (chirped) laser pulses for population transfer control. One such advancement is the C-STIRAP protocol \cite{Chirped_stirap}, wherein the system dynamics are controlled via carefully designed frequency-chirp profiles.
This approach proves especially effective in spectrally congested environments, where conventional adiabatic techniques often lose their selectivity.
Expanding upon this concept, \cite{Aleksanyan2021SOLO} developed a theoretical model for complete population transfer in a five-level
M-type system, based on linear frequency scanning of the laser field near the resonance conditions of each transition. This method enables sequential
and coherent excitation of level-to-level transitions and demonstrates high efficiency in managing the dynamics of multilevel quantum systems.
A comparative analysis with traditional STIRAP schemes revealed that the scanning-based approach offers notable advantages, both in terms of control
precision and numerical stability. Collectively, these studies underscore the considerable potential of frequency-controlled strategies in the domain
of coherent control for complex quantum architectures.

The development of coherent control methods for multilevel quantum systems has become one of the central directions in the field of quantum
technologies. In the work \cite{Optimal_pulse_sequences}, detailed strategies were proposed for optimizing pulse
sequences to enhance the efficiency of population transfer across multiple quantum levels. By employing numerical optimal control techniques,
the authors demonstrated that high-fidelity transfer can be achieved even under deviations from strict adiabatic conditions, provided that the
temporal profiles of the driving pulses are appropriately shaped. In contrast, the more recent study \cite{Ishkhanyan2024}
introduces an analytical reverse-engineering method for pulse design that yields a prescribed population dynamics in a nonlinear three-state
system with intermediate-level losses. This approach enables the minimization of loss-channel occupation, which is of critical importance for
applications sensitive to decoherence.

Concurrently, considerable effort has been directed towards accelerating adiabatic passage via so-called shortcuts to adiabaticity.
\cite{OptimalSTIRAP} developed a control scheme based on a spin-to-spring mapping, allowing for the
emulation of STIRAP-like dynamics over significantly reduced time intervals. This approach addresses one of the principal constraints of
conventional STIRAP—its reliance on long pulse durations—whilst retaining high fidelity in population transfer.

From the standpoint of practical application,\cite{Genov2023} demonstrated a robust two-level state-swapping
protocol inspired by STIRAP, reaffirming its utility in quantum information processing where reversibility and reliability are paramount.
Further broadening the scope, \cite{Singhal_2025} applied STIRAP-inspired logic to superconducting qubits,
illustrating how dual-rail quantum gates based on this principle can be realised with enhanced resilience to parameter noise—a promising step
towards fault-tolerant quantum logic in solid-state systems.

In this work, we develop a numerical approach to the problem of inverse quantum control, focusing on the automated design of external laser driving fields 
that enable high-precision and controllable population transfer in multi-level quantum systems. Our method is based on the optimization of control pulse parameters 
using gradient-based algorithms in combination with automatic differentiation, which enables efficient navigation of high-dimensional parameter spaces while maintaining 
the desired physical properties of the control fields. The inclusion of differentiable regularization terms—such as soft constraints on temporal ordering—ensures both 
control over pulse structure and smoothness of the loss landscape, thereby improving convergence stability.

We formulate a loss functional that accounts for both the dynamical behavior of the system and physical objectives for the final state, including the suppression of 
intermediate state populations and accurate targeting of the desired quantum configuration. The proposed framework is flexible and scalable, making it applicable to 
systems with arbitrary numbers of quantum levels, and especially valuable in situations where classical control methods, such as STIRAP, are inefficient or inapplicable.

In the sections that follow, we provide a detailed mathematical formulation, describe the architecture of the optimization strategy, and present numerical results that 
demonstrate the effectiveness and versatility of our approach. The outcomes of this work lay the groundwork for more general quantum control protocols aimed at 
experimental implementation in realistic physical platforms.

\section{Physical Background}

\begin{figure}[h]
	\centering
	\includegraphics[width=0.5\textwidth]{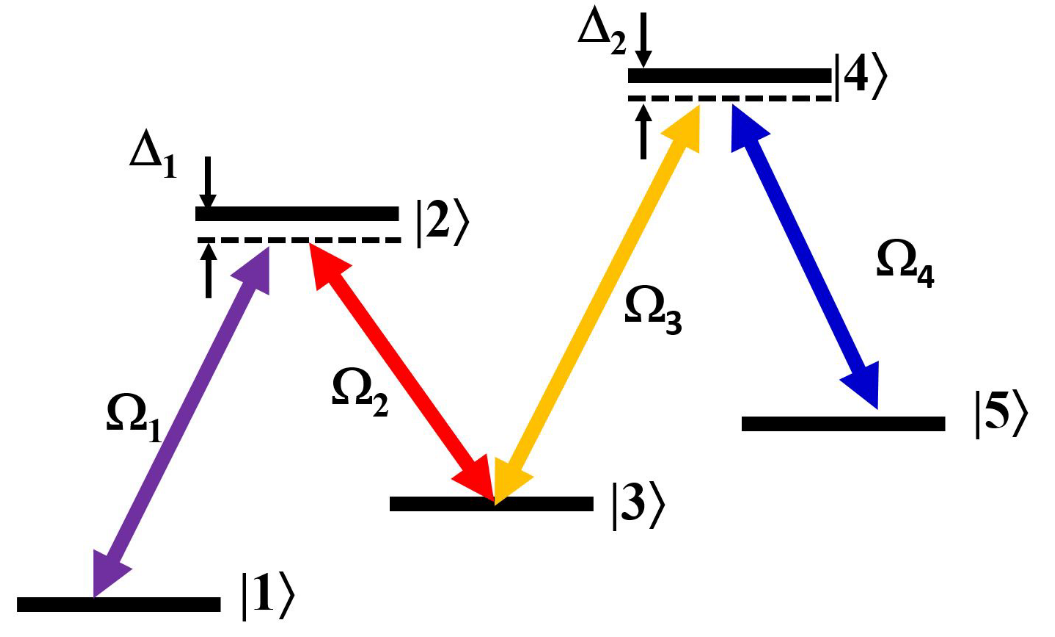}
	\caption{Schematic representation of the five-level M-type quantum system with four laser-driven transitions\cite{AllOpticalFourBitToffoliGate}. 
	The system consists of three ground states ($|1\rangle$, $|3\rangle$, $|5\rangle$) and two excited states ($|2\rangle$, $|4\rangle$). 
	The arrows indicate the dipole-allowed transitions induced by the laser fields.}
	\label{fig:5-level-system}
\end{figure}

We consider a five-level quantum system arranged in a so-called M-type configuration Fig.~\ref{fig:5-level-system},
consisting of three ground states $|1\rangle$, $|3\rangle$, and $|5\rangle$, and two excited
states $|2\rangle$ and $|4\rangle$. This system is driven by four coherent laser fields that induce
sequential dipole-allowed transitions among adjacent levels as illustrated by Eq.\eqref{eq:transitions}:

\begin{equation}
	\label{eq:transitions}
	|1\rangle \leftrightarrow |2\rangle,\quad
	|2\rangle \leftrightarrow |3\rangle,\quad
	|3\rangle \leftrightarrow |4\rangle,\quad
	|4\rangle \leftrightarrow |5\rangle.
\end{equation}

The time evolution of the density matrix of the system\cite{Blum_DM} $\rho \in \mathbb{C}^{5 \times 5}$ is governed by the Lindblad master equation\cite{LindbladMasterEquation},
which accounts for both coherent and incoherent processes:

\begin{equation}
	\frac{d\rho}{dt} = -i[H(t), \rho] + \mathcal{L}[\rho],
	\label{eq:lindblad}
\end{equation}

where $H(t)$ denotes the system Hamiltonian in the interaction picture and under the rotating-wave approximation,
while $\mathcal{L}[\rho]$
is the Lindblad superoperator describing dissipation.

In a five-level quantum system of M-type configuration, where sequential transitions between adjacent levels are allowed as 
\( |1\rangle \leftrightarrow |2\rangle \leftrightarrow |3\rangle \leftrightarrow |4\rangle \leftrightarrow |5\rangle \), 
the interaction Hamiltonian in the interaction picture under the rotating-wave approximation (RWA) is represented on the orthonormal 
basis \( \{ |1\rangle, |2\rangle, |3\rangle, |4\rangle, |5\rangle \} \) by the following matrix:
\begin{widetext}
\begin{equation}
	H(t) =
	\begin{pmatrix}
0 & \Omega_1(t) e^{-i\Delta_1 t} & 0 & 0 & 0 \\
\Omega_1(t) e^{i\Delta_1 t} & 0 & \Omega_2(t) e^{i\Delta_2 t} & 0 & 0 \\
0 & \Omega_2(t) e^{-i\Delta_2 t} & 0 & \Omega_3(t) e^{-i\Delta_3 t} & 0 \\
0 & 0 & \Omega_3(t) e^{i\Delta_3 t} & 0 & \Omega_4(t) e^{i\Delta_4 t} \\
0 & 0 & 0 & \Omega_4(t) e^{-i\Delta_4 t} & 0 \\
\end{pmatrix},
	\label{eq:hamiltonian}
\end{equation}
\end{widetext}
where \( \Omega_j(t) \) is the time-dependent Rabi frequency corresponding to the transition \( |j\rangle \leftrightarrow |j+1\rangle \), 
 characterizing the strength of the interaction between the atomic dipole and the external electromagnetic field, is given by
\[
\Omega = -\frac{\mathbf{d} \cdot \mathbf{E}_0}{\hbar}, 
\]
where \( \mathbf{d} \) is the electric dipole moment operator associated with the transition, \( \mathbf{E}_0 \) is the amplitude of the electric field. 

The driving field is assumed to be monochromatic and linearly polarized,
\( \mathbf{E}(t) = \mathbf{E}_0 \cos(\omega_L t) \), where \( \omega_L \) denotes the laser frequency. 
The detuning \( \Delta_j = \omega_{L,j} - \omega_{j+1,j} \) represents the difference between the frequency \( \omega_{L,j} \) of the laser 
field driving the transition \( |j\rangle \leftrightarrow |j+1\rangle \) and the resonant transition frequency \( \omega_{j+1,j} = (E_{j+1} - E_j)/\hbar \), 
which is determined by the energy spacing between the two atomic levels. This frequency mismatch plays a critical role in the excitation dynamics, especially under off-resonant driving, adiabatic regimes, and coherent population transfer protocols such as STIRAP, where precise control of the detuning determines the efficiency and robustness of quantum state manipulation.

The time dependence of \( \Omega_j(t) \) is typically realized through
Gaussian-shaped laser pulses and is parameterized as
\[
\Omega_j(t) = \Omega_{0,j} \exp\left( -\frac{(t - t_{0,j})^2}{\sigma_j^2} \right),
\]
where \( \Omega_{0,j} \) is the peak Rabi frequency, \( t_{0,j} \) is the pulse center in time, and \( \sigma_j \) denotes the temporal width. This framework 
enables efficient coherent control of quantum state dynamics in complex M-type multilevel systems.

The dissipation is described by the Lindblad superoperator in Eq.\eqref{eq:lindblad_operator}:

\begin{equation}
	\mathcal{L}[\rho] = \sum_k \gamma_k \left( L_k \rho L_k^\dagger - \frac{1}{2} \left\{ L_k^\dagger L_k, \rho \right\} \right),
	\label{eq:lindblad_operator}
\end{equation}

where $L_k$ are the collapse operators and $\gamma_k$ are their respective decay rates. For the M-type system\cite{Pogosyan2023}, we consider the
representative  decay channels listed in Eq.\eqref{eq:jump_operators}:
\begin{widetext}
\begin{equation}
	L_1 = |1\rangle\langle2|,\quad
	L_2 = |3\rangle\langle2|,\quad
	L_3 = |3\rangle\langle4|,\quad
	L_4 = |5\rangle\langle4|.
	\label{eq:jump_operators}
\end{equation}
\end{widetext}
The decay processes are characterized by two distinct rates. The quantity $\Gamma$ denotes the natural (longitudinal)
decay rate of the excited states. The transverse decay rate $\gamma$ governs the dephasing of coherences and includes both spontaneous
emission and additional homogeneous broadening, as given in Eq.~\eqref{eq:gamma}:

\begin{equation}
	\gamma = \frac{\Gamma}{2} + \gamma_c,
	\label{eq:gamma}
\end{equation}

where $\gamma_c$ is a phenomenological\cite{Steck2007} parameter accounting for collisional dephasing or other environmental effects.

The theoretical framework presented above offers a complete and rigorous model for analyzing coherent control and dissipative dynamics in multilevel M-type quantum systems. 
It supports detailed numerical studies aimed at optimizing population transfer from $|1\rangle$ to $|5\rangle$ with minimal occupation of the lossy intermediate states 
$|2\rangle$ and $|4\rangle$, under realistic conditions that include detuning, pulse shaping, and decoherence.

\section{Mathematical Formulation and Parameter Estimation}

In this section, we describe the numerical strategies used to estimate the parameters of our system of ordinary differential equations (ODEs)\cite{Ixaru2004_RK_ODE_Methods}. 
We leverage the Julia \texttt{Optimization.jl}\cite{vaibhav_kumar_dixit_2023_7738525} framework in combination with automatic differentiation and gradient-based 
solvers to minimize a customized loss function.

For the numerical integration of the ODEs required in our evaluation of the loss function, we employ the 
\texttt{DifferentialEquations.jl}\cite{rackauckas2017differentialequations} package. This sophisticated suite provides a 
comprehensive collection of solvers customized for various characteristics of problems. We primarily utilize the Tsitouras 5/4 Runge-Kutta method with 
adaptive time-stepping for its excellent balance of accuracy and computational efficiency. The package's automatic differentiation compatibility 
enables efficient gradient computation through the ODE solutions, which proves crucial for accurate sensitivity analysis and gradient-based parameter optimization. Evaluation of each parameter set requires solving the complete system of equations to generate the model trajectories that are then compared against experimental data in our loss function.

\subsection{Mathematical Formulation}

Let $\mathbf{p} \in R^m$ denote the vector of unknown parameters and consider the initial value problem:

\begin{equation}
\dot{\rho} = M\bigl(\mathbf{\rho}(t), t,\mathbf{p}\bigr), \quad \mathbf{\rho}(0) = \mathbf{u}_0, \quad t\in[0,T].
\end{equation}

We define a loss functional $\mathcal{L}(\mathbf{p})$ that measures the discrepancy between simulated trajectories and desired behaviors:

\begin{equation}
\mathcal{L}(\mathbf{p}) = \int_{\frac{T}{2}}^T \rho_{11}(t)\,dt \;+\;\sum_{k = 2}^4\left[\int_0^T \rho_{kk}(t)\,dt\right] \;+\;\bigl[\rho_{55}(T)-1\bigr]^2,
\end{equation}

The first term restricts the accumulation of population in the initial state $|1\rangle$ to the latter half of the protocol, discouraging premature transfer. 
The second term penalizes populations in intermediate states $|2\rangle$, $|3\rangle$, and $|4\rangle$ throughout the control interval. 
The final squared-error term drives the population of target state $|5\rangle$ to unity at the terminal time, ensuring high-fidelity transfer.

\subsection{Regularization and Constraints}

We solve the inverse control problem of finding pulse parameters that maximize coherent transfer ($|1\rangle\to|5\rangle$) while suppressing lossy intermediate populations. This is cast as a nonlinear, bound-constrained optimization of the Gaussian-pulse parameters:

\begin{equation}
\{\,t_i,\sigma_i,\mathcal{E}_i^{(0)},\Delta_i\}_{i=1}^4\;,
\end{equation}

with physically motivated lower and upper bounds:

\begin{equation}
\begin{cases}
t_i \in [15, 35] \\ 
\sigma_i \in [2, 4] \\
\mathcal{E}_i^{(0)} \in [1, 35] \\
\Delta_i \in [-5, 5]
\end{cases}
\end{equation}

To guide the optimization toward physically meaningful regimes, we incorporate both hard-box constraints and soft penalties. Soft constraints include smooth 
penalties to prevent parameters from drifting outside experimentally viable ranges, implemented using softplus-based barriers.

\subsection{Generalized Temporal Ordering Regularization}

To guide the optimizer toward or away from specific pulse orderings, let \(\mathbf{t} = (t_1, t_2, t_3, t_4)\) be the vector of pulse centers 
and choose a reference index \(j \in \{1,2,3,4\}\). We introduce the following differentiable ordering penalty:
\begin{widetext}
\begin{equation} 
  P(\mathbf{t}; j, s, k_\text{sharp})
  = \prod_{\substack{k=1 \\ k \ne j}}^4
    \sigma\bigl(s\,k_\text{sharp}\,(t_j - t_k)\bigr),
  \quad
  \sigma(x) = \frac{1}{1 + e^{-x}},
\end{equation}
\end{widetext}
where:
\begin{itemize}
  \item \(k_\text{sharp}>0\) controls the steepness of each sigmoid transition,
  \item \(s\in\{+1,-1\}\) encodes the desired ordering:
    \begin{enumerate}
      \item \(s=+1\): enforce \(t_j > t_k\) (\(j\) occurs \emph{last}),
      \item \(s=-1\): enforce \(t_j < t_k\) (\(j\) occurs \emph{first}).
    \end{enumerate}
\end{itemize}

Each term \(\sigma\bigl(s\,k_\text{sharp}(t_j - t_k)\bigr)\) satisfies
\begin{widetext}
\[
  \sigma\bigl(s\,k_\text{sharp}(t_j - t_k)\bigr)
  \approx
  \begin{cases}
    1, & \text{if the desired ordering holds (e.g.\ }t_j \gg t_k\text{)},\\
    0, & \text{if it is strongly violated (e.g.\ }t_j \ll t_k\text{)}.
  \end{cases}
\]
\end{widetext}
Because \(\sigma(x)\) is smooth, this penalty remains differentiable, yet becomes sharply selective as \(k_\text{sharp}\) grows. 
By multiplying over all \(k\neq j\), \(P\) softly—but effectively—encourages a global ordering of pulse \(j\) relative to the others, 
while still allowing the optimizer to explore near-boundary regions of parameter space.

\medskip

\noindent\textbf{Chained Ordering Extension.} One can further enforce complex sequences by chaining pairwise comparisons in series.  For example, to require
\[
  t_4 > t_2 > t_3 > t_1,
\]
define three sigmoid factors
\(\sigma\bigl(k_\text{sharp}(t_4 - t_2)\bigr)\),
\(\sigma\bigl(k_\text{sharp}(t_2 - t_3)\bigr)\),
\(\sigma\bigl(k_\text{sharp}(t_3 - t_1)\bigr)\),
and form their product.  This “chain of order” acts like a single soft indicator of the entire pattern: it approaches one only 
when each successive inequality holds, and decays smoothly otherwise.  Such a construction lets one encode arbitrary multistep 
temporal patterns, all within a fully differentiable framework.

Such generalized temporal‐ordering regularization enhances control flexibility and supports protocol discovery, especially 
in systems where canonical STIRAP-like pulse sequences are suboptimal or infeasible.

\subsection{Optimization Strategy and Implementation}

\subsubsection{Automatic Differentiation}
We initially implemented gradient computations via forward-mode finite differences using \texttt{AutoFiniteDiff} (backed by FiniteDiff.jl), 
which provided a robust baseline and useful cross-checks.  We then transitioned to source-to-source reverse-mode automatic differentiation with Zygote.jl 
\cite{Zygote.jl-2018} (via the \texttt{AutoZygote} interface) to obtain exact gradients up to machine precision and better performance.  This AD approach 
scales efficiently to high-dimensional parameter vectors and integrates seamlessly with our Julia-based simulation and loss pipelines.  As a fallback—used 
in rare cases where reverse-mode AD is impractical due to memory constraints or non-differentiable components—we still retain the finite-difference implementation.  
Though each finite-difference gradient requires \(n+1\) function evaluations for \(n\) parameters, it remains valuable for debugging and small-scale tests.

\subsubsection{Optimization Algorithms}
After benchmarking a variety of solvers, we selected the Limited-memory BFGS (L-BFGS) family for its blend of performance and convergence quality. The standard L-BFGS 
\cite{10.1145/279232.279236} algorithm (unconstrained)  approximates the inverse Hessian by storing only the most recent \(m\) vector pairs \(\{s_i,y_i\}\), 
reducing both storage and per-iteration cost to \(\mathcal{O}(mn)\), where \(n\) represents the number of parameters. For handling box constraints, L-BFGS-B
 (bound constrained) \cite{Liu1989} extends the base algorithm to accommodate constraints of the form \(l_i \le x_i \le u_i, \quad i=1,\dots,n\), as specified in
  Eq.\,(13). At each step, this variant computes a projected gradient ("generalized Cauchy point") within these bounds, and minimizes a 
  limited-memory Hessian model under the feasible region.

To validate robustness, we also tested several alternative approaches. Our evaluation included stochastic and derivative-free methods such as Particle Swarm 
Optimization and SAMIN, as well as deterministic approaches including Newton–Raphson and NGMRES. While none of these alternatives surpassed L-BFGS in terms of 
convergence speed or final loss values, SAMIN demonstrated particular effectiveness at escaping local minima in noisy landscapes.

\subsubsection{Implementation and Diagnostics}
All solvers and differentiation backends are accessed via the Optimization.jl ecosystem, with callback hooks configured for real-time monitoring.  
These diagnostics—implemented through user callbacks—provide interactive loss and gradient visualizations, enabling rapid identification of convergence 
issues and on-the-fly parameter steering.


\section{Results}

\begin{widetext}

\begin{figure}[h!] 
    \centering
    \includegraphics[width=1\textwidth]{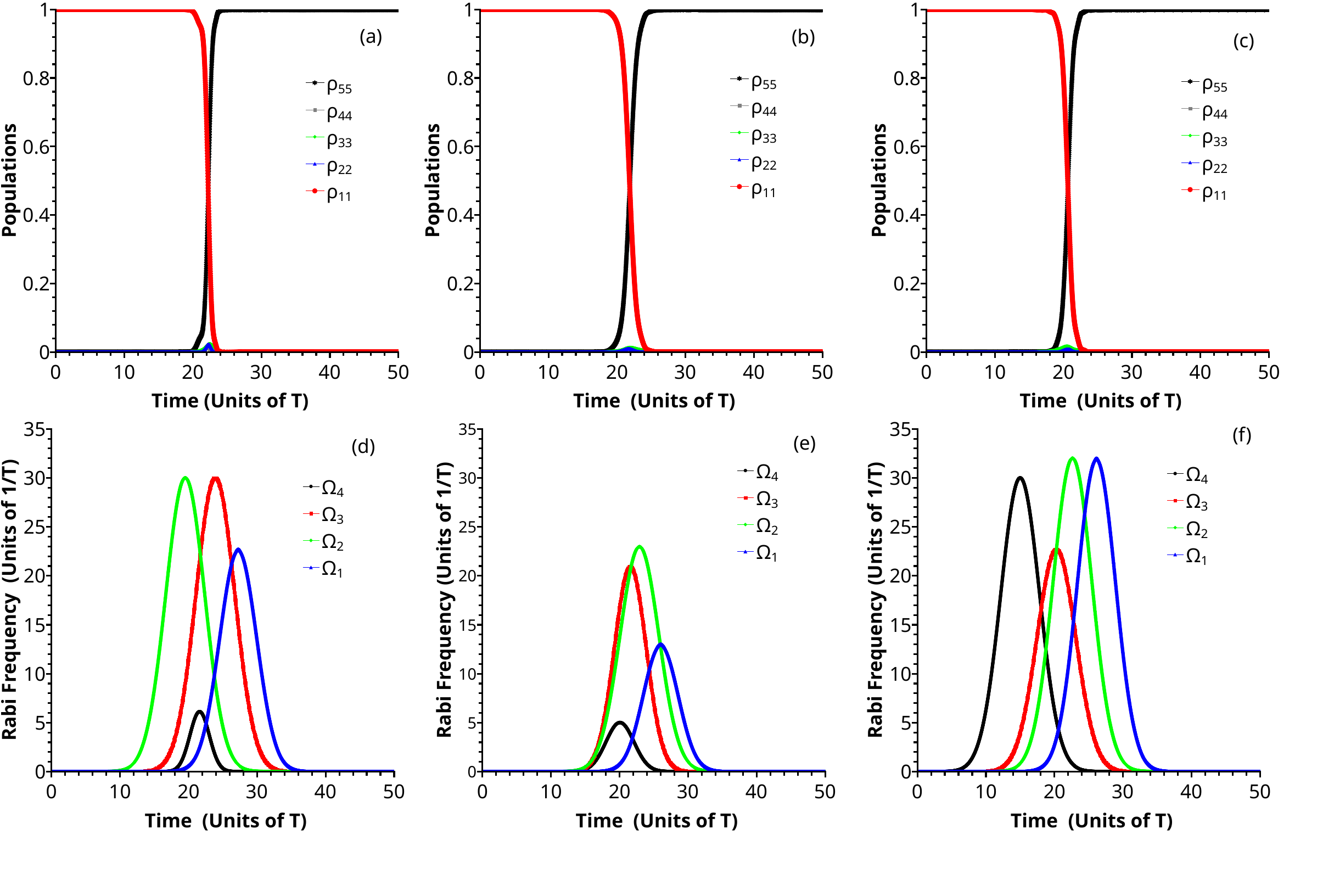}
    \caption{Time evolution of populations $\rho_{ii}(t)$ for $i = 1,\dots,5$ and corresponding Rabi frequencies $\Omega_i(t)$ ($i = 1,\dots,4$) 
    in an optimized M-type five-level quantum system. T is the duration of the shortest of pulses.}
    \label{fig:population-dynamics}    
\end{figure}
\end{widetext}

The Fig.\ref{fig:population-dynamics} visualize the evolution of a five-level quantum system arranged in an M-type configuration\cite{Gazazyan2017,Gazazyan2018}, 
in which three states are ground levels ($\ket{1}$, $\ket{3}$, $\ket{5}$) and two are excited states ($\ket{2}$ and $\ket{4}$). 
The system dynamics are governed by interactions with four coherent laser fields, each inducing a dipole-allowed transition between 
neighboring states: $\ket{1} \leftrightarrow \ket{2}$, $\ket{2} \leftrightarrow \ket{3}$, $\ket{3} \leftrightarrow \ket{4}$, 
and $\ket{4} \leftrightarrow \ket{5}$. Control over this multilevel transition is implemented via time-dependent Gaussian-shaped laser pulses, 
whose parameters (amplitudes, centers, and widths) are numerically optimized using gradient-based methods supported by automatic differentiation.

In the present model, effects related to atomic motion are neglected, including both thermal translational motion and possible Doppler shifts 
that could affect the resonance conditions of interaction with the laser pulses. Assuming purely radiative damping, the pure dephasing rate is 
taken as $ \gamma = \Gamma / 2 $, such that the excited-state populations relax toward steady-state values governed by the competition between 
coherent driving and spontaneous emission. Throughout all simulations, we employ a dimensionless formulation normalized to the natural decay 
rate of the excited states. A physical value of $\Gamma = 5$ MHz is chosen. Accordingly, time is measured in units of $1/\Gamma = 0.2,\mu\text{s}$, and all frequency-related 
parameters—including Rabi frequencies, detunings, and dephasing rates—are expressed in units of $\Gamma$.

The parameters used for the numerical modeling are presented in the table and correspond to four 
interaction channels within the quantum system. Each channel is characterized by four key quantities: the pulse peak 
time \( t_0 \), its width \( \sigma_0 \), the Rabi frequency amplitude \( \Omega_0 \), and the detuning \( \Delta \). The numerical values are listed below:

\begin{table}[h!]
\centering
\begin{tabular}{|c|c|c|c|c|}
\hline
\# & $t_0$ [$1/\Gamma$] & $\sigma_0$ [$1/\Gamma$] & $\Omega_0$ [$\Gamma$] & $\Delta$ [$\Gamma$] \\

\hline
1 & 18.9032811 & 3.227598027 & 3.552287843 & -0.208536178 \\
2 & 14.95405537 & 4.942187821 & 28.22099145 & 4.953080222 \\
3 & 20.94907939 & 5.234351941 & 33.03312243 & 0.061713435 \\
4 & 16.89870222 & 3.616342008 & 3.472354099 & -5.125913793 \\
\hline
\end{tabular}
\caption{Optimized pulse parameters for the control protocol shown in Figure 2d. 
Each pulse is defined by its temporal center $t_0$, width $\sigma_0$, amplitude $\Omega_0$, and detuning $\Delta$. 
The parameters were obtained via constrained optimization to ensure adiabatic and efficient population transfer in a multilevel quantum system.}
\label{tab:fig2d}
\end{table}

\begin{table}[h!]
\centering
\begin{tabular}{|c|c|c|c|c|}
\hline
\# & $t_0$ [$1/\Gamma$] & $\sigma_0$ [$1/\Gamma$] & $\Omega_0$ [$\Gamma$] & $\Delta$ [$\Gamma$] \\

\hline
1 & 25.99837314 & 3.622208677 & 13.00608085 & 3.571861571 \\
2 & 22.90506102 & 4.021947620 & 22.96440215 & 3.574703202 \\
3 & 21.56536551 & 3.282861957 & 20.91424811 & -3.255710582 \\
4 & 20.05748673 & 2.888952454 & 5.013657599 & -3.061313302 \\
\hline
\end{tabular}
\caption{Pulse parameters corresponding to Figure 2e.}
\label{tab:fig2e}
\end{table}

\begin{table}[h!]
\centering
\begin{tabular}{|c|c|c|c|c|}
\hline
\# & $t_0$ [$1/\Gamma$] & $\sigma_0$ [$1/\Gamma$] & $\Omega_0$ [$\Gamma$] & $\Delta$ [$\Gamma$] \\

\hline
1 & 26.13033743 & 3.940057889 & 32.00000000 & 0.101091573 \\
2 & 22.63505033 & 4.000000000 & 32.00000000 & 0.000448309 \\
3 & 20.28044792 & 4.000000000 & 22.68347007 & -0.004890434 \\
4 & 15.00000000 & 4.000000000 & 30.00000000 & -0.003750851 \\
\hline
\end{tabular}
\caption{Pulse parameters corresponding to Figure 2f.}
\label{tab:fig2f}
\end{table}

These parameters were obtained through numerical optimization and ensure efficient population transfer in the multilevel quantum system. 
They minimize the transient occupation of intermediate excited states while maintaining the required adiabaticity conditions.

The upper panels of the plots display the diagonal elements of the density matrix, $\rho_{11}(t)$ through $\rho_{55}(t)$, representing the 
time evolution of the probability of the system being in each corresponding state. The initial condition is such that the entire population 
resides in the ground state $\ket{1}$. As the sequence of pulses progresses, $\rho_{11}(t)$ decreases, reflecting depopulation of state $\ket{1}$, 
and population is gradually transferred to other states. The behavior of the excited states $\ket{2}$ and $\ket{4}$ is particularly critical, 
as they are subject to dissipation via spontaneous decay, which is incorporated through the Lindblad formalism with collapse operators. The 
plots clearly show that $\rho_{22}$ and $\rho_{44}$ remain close to zero, indicating that the system avoids real occupation of these levels and 
instead passes through them virtually, via quantum interference — a key mechanism in adiabatic population transfer of the STIRAP type\cite{AllOpticalFourBitToffoliGate}. 
The intermediate state $\ket{3}$ may become transiently populated, but its occupation is also suppressed through the pulse optimization process. At the final time, the 
curve $\rho_{55}(t)$ reaches a value close to unity, signifying nearly complete population transfer to the target state $\ket{5}$, which is the main indicator 
of protocol efficiency.

The lower panels illustrate the time profiles of the four Rabi frequencies $\Omega_1(t)$ through $\Omega_4(t)$, each associated with a specific transition 
between adjacent levels. These profiles are modeled as Gaussian pulses characterized by amplitude, center, and width. These pulse parameters are the subject 
of optimization with the objective of minimizing dissipation, maximizing final population in state $\ket{5}$, and enforcing temporal ordering constraints. 
The configuration of the pulses is chosen to prevent premature activation of transitions and to enable coherent population transfer along the state chain. 
Often, a counter-intuitive pulse sequence is used (e.g., activating $\Omega_4$ before $\Omega_1$), consistent with stimulated Raman adiabatic passage (STIRAP) techniques, 
allowing the system to bypass real excitation of lossy intermediate states. Visually, this appears as temporally overlapped pulses with specific delays, which are 
critical for the correct sequencing of transitions. The shape of the curves $\Omega_i(t)$ illustrates how the laser fields are turned on and off in time, structuring 
the quantum interference pathways of system evolution.

Each pair of graphs corresponds to a different stage of the numerical experiment—namely, the initial configuration, intermediate optimization steps, and the final 
optimized result. In all cases, the characteristic dynamics are clearly visible: smooth pulse shaping, precise population transfer, and effective suppression of 
unwanted intermediate-state occupations. These plots represent the direct numerical integration of a system of ordinary differential equations, incorporating both 
a time-dependent Hamiltonian and dissipative processes modeled via the Lindblad superoperator. Implementing such a numerical protocol requires high precision in 
both the ODE solver and the gradient-based parameter optimization, which is achieved using specialized Julia libraries such as \texttt{DifferentialEquations.jl} 
and \texttt{Optimization.jl}. The resulting plots confirm that the optimized control parameters enable nearly perfect population transfer with minimal loss, making 
the proposed scheme highly promising for both theoretical investigations and experimental realizations in quantum systems.

\section{Conclusion}

We present a numerical optimization framework for enhancing population transfer in multilevel quantum systems. The approach proves 
particularly effective for implementing Stimulated Raman Adiabatic Passage (STIRAP) in five-level M-type systems, which serves as 
a fundamental protocol for coherent quantum state manipulation. Our method integrates automatic differentiation,adaptive differential equation solvers, and constrained 
gradient-based optimization, employing the L-BFGS-B algorithm to identify physically realizable solutions within strict physical boundaries efficiently.

Extensive numerical experiments demonstrate that the proposed strategy is highly flexible and robust.
It enables high-fidelity population transfer from the initial to the target state, minimizing transient occupation of
intermediate levels while generating experimentally feasible control pulses. The examples presented highlight only
a subset of the dynamical scenarios achievable by this approach. The methodology is broadly applicable to a range
of quantum control tasks, including accelerated adiabatic protocols, quantum memory writing, and the design of
 quantum gates\cite{Aleksanyan2018,Aleksanyan2021SOLO,Gazazyan2009,AllOpticalFourBitToffoliGate}.

Overall, the combination of modern optimization techniques with precise quantum simulations provides a
powerful platform for coherent control in complex quantum architectures, paving the way toward more flexible,
scalable, and resilient quantum technologies.

\section{Acknowledgement}

The work was supported by the Higher Education and Science Committee of RA (project No. 1-6/IPR) and ().
We thank the Armenian National Supercomputing Center(ANSCC)\cite{anscc2025} for providing the essential resources and support
that made this research possible.

\bibliographystyle{ieeetr}
\bibliography{mycite}

\end{document}